\documentstyle[prl,aps,twocolumn,amssymb,graphicx]{revtex}
\def\bea{\begin{eqnarray}}
\def\eea{\end{eqnarray}}
\def\nn{\nonumber}
\renewcommand\epsilon{\varepsilon}
\def\beq{\begin{equation}}
\def\eeq{\end{equation}}

\def\lsim{\mathrel{\raise.3ex\hbox{$<$\kern-.75em\lower1ex\hbox{$\sim$}}} }
\def\gsim{\mathrel{\raise.3ex\hbox{$>$\kern-.75em\lower1ex\hbox{$\sim$}}} }

\begin{document}
\draft
\title{CP Violation in neutrino oscillation and leptogenesis}
\author{T. Endoh$^{(a)}$ \thanks{E-mail:endoh@theo.phys.hiroshima-u.ac.jp},
S. Kaneko$^{(b)}$ \thanks{E-mail:kaneko@muse.sc.niigata-u.ac.jp},
S. K. Kang$^{(a)(c)}$
\thanks{
E-mail:kang@theo.phys.hiroshima-u.ac.jp},
T. Morozumi$^{(a)}$ \thanks{E-mail:morozumi@theo.phys.sci.hiroshima-u.ac.jp},
and M. Tanimoto$^{(b)}$ \thanks{E-mail:tanimoto@muse.sc.niigata-u.ac.jp}\\
\small   $^{(a)}$Graduate School of Science, Hiroshima University,
         Higashi Hiroshima -739-8526, Japan\\
\small $^{(b)}$Department of Physics, Niigata University, Niigata
- 950-2181, Japan \\
\small $^{(c)}$Department of Physics, Seoul National University,
Seoul 151-747, Korea}
\maketitle
\thispagestyle{empty}
\begin{abstract}
We study the correlation between CP violation in neutrino
oscillations and leptogenesis in the framework with two heavy
Majorana neutrinos and three light neutrinos. Among three
unremovable CP phases, a heavy Majorana phase contributes to
leptogenesis. We show how the heavy Majorana phase contributes to
Jarlskog determinant $J$ as well as neutrinoless double $\beta$
decay by identifying a low energy CP violating phase which signals
the CP violating phase for leptogenesis. For some specific cases
of the Dirac mass term of neutrinos, a direct relation between
lepton number asymmetry and $J$ is obtained. For the most general
case of the framework, we study the effect on $ J $ coming from
the phases which are not related to leptogenesis, and also show
how the correlation can be lost in the presence of those phases.
\end{abstract}
 \narrowtext
Finding any relation between  baryogenesis via leptogenesis
\cite{Fu} and low energy CP violation observed in the laboratory
is a very interesting issue \cite{cpbary}. The CP violation
required for leptogenesis stems from the CP phases in the heavy
Majorana sector, whereas CP violation measurable from the neutrino
oscillations \cite{necp} can be described by the neutrino mixing
matrix. One interesting question concerned with the low energy
leptonic CP violation is whether it can be affected by the CP
violating phases responsible for leptogenesis. Several people
\cite{leptcp} have already discussed some potential connections
between low energy CP violation and leptogenesis by using some
ansatz, but it is still unclear how large the former can affect
the latter in general. The major difficulty to quantify such a
connection occurs due to lack of the available low energy data to
fix parameters of the seesaw model.

The purpose of this paper is to examine in a rather general
framework how leptogenesis can be related to the low energy CP
violation by determining the parameters as many as possible from
available low energy experimental results and cosmological
observations. In order to make a quantitative analysis of the
connection between low energy leptonic CP violation and
leptogenesis, we consider the {\it minimal} CP violating seesaw
model which has two heavy Majorana neutrinos and three light
left-handed neutrinos; (3,2) seesaw model. As will be shown later,
to break CP symmetry, the required minimal number of singlet heavy
Majorana neutrino is two in the seesaw model with three light
lepton doublets. This (3,2) seesaw model is consistent with recent
data of neutrino oscillations and contains 8 real parameters and 3
CP violating phases in the neutrino sectors which make this model
more constrained and predictive compared with the general (3,3)
seesaw model \cite{YaGe} with 18 parameters. We will show that
while all three CP violating phases contribute to low energy
leptonic CP violation, only a single CP violating phase
contributes to leptogenesis. We will also investigate how large
the CP phase responsible for leptogenesis contributes to low
energy CP violation  by determining the independent parameters
from available experimental results and cosmological observations.
Finally, we will discuss the potential implication of CP violation
measurable from neutrino oscillations on leptogenesis.

Let us begin our study by considering the leptonic sector of the
(3,2) seesaw model. In a basis where both heavy Majorana and
charged lepton mass matrices are real diagonal, the Lagrangian is
given by: \bea {\cal L}= -\overline{l_{iL}} {m_l}_i l_{iR} -
\overline{\nu_{Li}} {m_D}_{ij} N_{Rj} - \frac{1}{2}
\overline{(N_{Rj})^c} M_j N_{Rj}, \label{lag} \eea where
$i=1,2,3,~ j=1,2$ and the Dirac mass term $m_D$ is $ 3 \times 2$
matrix. Here, we remark that the Dirac mass matrix $m_D$ contains
$3N-3$ unremovable CP phases if we take $N$ singlet heavy Majorana
neutrinos in this basis. Thus, one can easily see that at least
two singlet heavy Majorana neutrinos are required to break CP
symmetry in the seesaw model with three lepton $SU(2)$ doublets.
The $3\times 2$  matrix $m_D$ can be generally parameterized as:
\bea m_D =U_L \left( \begin{array}{cc}
    0 & 0 \\
    m_{2} & 0 \\
    0 & m_{3}
    \end{array} \right) V_R, \label{mD}
\eea with, \bea U_L&=&
   O_{23}(\theta_{L23})
   U_{13}(\theta_{L13},\delta_L) O_{12}(\theta_{L12})
   P(\gamma_L) , \nn \\
V_R&=&\left(\begin{array}{cc}
      \cos\theta_R  & \sin\theta_R  \\
      -\sin\theta_R  & \cos\theta_R \end{array} \right)
      \left(\begin{array}{cc}
      e^{-i \frac{\gamma_R}{2}} & 0 \\
      0 & e^{i  \frac{\gamma_R}{2}}
      \end{array} \right),
\label{mD} \eea
 where $O_{ij}$ and $U_{ij}$ denote the rotations of $(i,j)$
 plane,
$P(\gamma_L)=diag.[1,e^{-i\gamma_L/2}, e^{i\gamma_L/2}]$, and
$m_2$ and $m_3 $ are real and positive. Without loss of
generality, we can choose $ m_2 \le m_3 $. The allowed range for
the angles and the phases is  $ [-\pi ,\pi] $.
There are three CP violating phases, $\gamma_R$ which appears in
$V_R$, $\gamma_L$ and $\delta_L$ in $U_L$. In a different basis
with complex $M_i$, $\gamma_R$ can be interpreted as a heavy
Majorana phase.
The lepton number
asymmetry for the lightest heavy Majorana neutrino ($N_1$) decays
into $l^{\mp} \phi^{\pm}$ \cite{ep} is given by;
 \bea
\epsilon_1 =
\frac{\Gamma_1-\overline{\Gamma_1}}{\Gamma_1+\overline{\Gamma_1}}
=
  -\frac{3 M_1}{2 M_2 V^2}
  \frac{Im[\{({m_D}^{\dagger} m_D)_{12}\}^2]}
{({m_D}^{\dagger} m_D)_{11}}, \eea where $V=\sqrt{4 \pi} v$ with
$v=246$ GeV, and \bea Im[\{({m_D}^{\dagger}m_D)_{12}\}^2]=
\left(\frac{m_2^2-m_3^2}{2}\right )^2 \sin^2 2\theta_R  \sin 2
\gamma_R. \label{ep1} \eea We see that CP violation concerned with
leptogenesis can be possible only if the mixing angle $\theta_R$
and CP violating phases $\gamma_R$ for the heavy Majorana
neutrinos are non-zero.
For our purpose, let us now study how the phase
$\gamma_R$ contributes to CP violation in the neutrino
oscillations which is usually described in terms of the MNS
neutrino mixing matrix \cite{MaSa}. The effective mass matrix for
light neutrinos is given by $ m_{eff}=-{m_D}\frac{1}{M}
{m_D}^T=-U_L m V_R \frac{1}{M} {V_R}^T m^T U_L^T \label{meff} $,
and is diagonalized by the MNS mixing matrix as $
U_{MNS}^{\dagger} m_{eff} U_{MNS}^{\ast}= diag[n_1,n_2,n_3]$.
Then, the MNS mixing matrix  is decomposed into two mixing
matrices as follows;
 \bea
 U_{MNS}= U_L K_R, \label{MNS}
 \eea
 where
$K_R$ is a unitary  matrix diagonalizing the matrix
 $Z\equiv -m V_R \frac{1}{M} V_R^T m^T$ and  parameterized by,
\bea K_R =\left(\begin{array}{ccc}
 1 & 0 & 0 \\
 0 & \cos\theta & \sin\theta e^{-i \phi} \\
 0 & -\sin \theta e^{i \phi}  & \cos\theta
      \end{array} \right)
      \left(\begin{array}{ccc}
      1 & 0 & 0 \\
      0 & e^{i \alpha} & 0 \\
      0 & 0 & e^{-i \alpha} \\
      \end{array} \right).
\label{K} \eea Then, $ -K_R^{\dagger} m V_R \frac{1}{M} V_R^T m^T
K_R^{\ast}= diag.[0, n_2, n_3]$. Note that the (3,2) seesaw model
predicts one massless neutrino. In addition, $\theta$, $\phi$ and
$\alpha$ are determined as: \bea
&& \phi=Arg.(Z_{22}^{\ast} Z_{23}+ Z_{23}^{\ast} Z_{33}), \nn \\
&& \tan 2\theta= \frac{2 |Z_{22}^{\ast} Z_{23}+ Z_{23}^{\ast}
Z_{33}|}
                 {|Z_{33}|^2-|Z_{22}|^2}, \\
&&  2\alpha =Arg.[\cos^2\theta Z_{22}+
 \sin^2 \theta Z_{33} e^{- 2i \phi} -\sin2\theta Z_{23}e^{-i \phi}].\nn
\label{phi} \eea We remark that the mixing angle $\theta_R$ and CP
violating phase $\gamma_R$ have been transferred to $\theta, \phi$
and $\alpha$. As one can see from the above formulae, leptogenesis
occurs only if the mixing angle $\theta_R $ and   CP violating
phase $ \gamma_R$ are non-zero, which in turn implies
non-vanishing
 $\phi$, $\alpha$, $\theta$ via Eq.(8).
 As we will see below, the CP phase $\phi$ contributes to
CP violation in neutrino oscillations, so that it is anticipated
that there is correlation between CP violation generated from
neutrino mixings and leptogenesis. To see this concretely, let us
compute Jarlskog determinant \cite{Jals} $J=Im [U_{MNS e 1} U_{MNS
e 2}^{\ast} U_{MNS \mu 1}^{\ast}  U_{MNS \mu 2}],$ which is
proportional to the CP asymmetry in neutrino oscillation, $\Delta
P=P_{(\nu_{\mu} \rightarrow\nu_{e})} -{\bar P}_{(\bar{\nu_{\mu}}
\rightarrow \bar{\nu_e})} =4 J \left(\sin(\frac{\Delta m_{12}^2
L}{2E})+
                  \sin(\frac{\Delta m_{23}^2 L}{2E})+
                  \sin(\frac{\Delta m_{31}^2 L}{2E}) \right) $.
By using Eq.(6), we obtain: \bea J&=&\frac{1}{8} \sin 2
\theta_{L12} \sin 2 \theta_{L13}
  [c_{L13} \cos 2 {\theta}  \sin \delta_L \sin 2 \theta_{L23}   \nn \\
 &+& c_{L12} \sin 2 {\theta}
  \sin (\delta_L -\gamma_L -{\phi}) \cos 2 \theta_{L23}  \nn \\
  &-& \frac{1}{2} s_{L12} s_{L13}
 \sin 2 {{\theta}} \sin 2 \theta_{L23} \sin(2 \delta_L-\gamma_L -
 {\phi})]   \nn  \\
  &+& \frac{1}{8} \sin 2 {{\theta}} \sin 2 \theta_{L23} \sin(\gamma_L +
  {{\phi}}) \times \nn \\
  && (\sin 2 {\theta}_{L12} c_{L13} s_{L12}-
             \sin 2 {\theta}_{L13} s_{L13} c_{L12})
\label{jarls}. \eea
 From the expression of $J$, it is obvious that all three CP
violating phases $\delta_L, \gamma_L$ and $\phi$ contribute to CP
violation in the neutrino oscillations, and that the CP phase
$\gamma_L$ always hangs around $\phi$. Since only $\phi$ is
closely related to leptogenesis, in order to investigate the
interplay between CP violation for leptogenesis and low energy
leptonic CP violation, we should determine  the contributions of
the phases $(\delta_L,\gamma_L$) and $\phi$ separately as well as
to fix the parameters $\theta's$.

Before discussing the correlation between both CP violations, let
us study how
 we can get some information on the mixing angles and CP phases from
 the available experimental and cosmological results.
The mixing angles and CP phases can be classified into two
categories, one contains $\theta, \phi$ and $\alpha$ which are
related to phenomena at high energy and the other contains
parameters in $U_L$. First of all, we show how we can estimate the
allowed values of CP violating phase $\phi$ and mixing angle
$\theta$. The information on $\phi$ and $\alpha$ may come from the
constraints on light neutrino mass spectra as well as cosmological
condition for leptogenesis. To see this, we first present the
parameters $\gamma_R, \theta_R, m_2, m_3$ and lepton number
asymmetry $\epsilon_1$ in terms of some physical quantities which
will be taken as inputs in numerical calculation. Here, we choose
the heavy Majorana neutrino masses ($M_1, M_2$), their decay
widths ($\Gamma_1, \Gamma_2$), and light neutrino masses
$(n_2,n_3)$ as the physical input parameters. As will be clear
later, it is convenient to define two parameters $x_i (i=1,2)$;
 \bea
 x_i=\frac{({m_D}^{\dagger} m_D)_{ii}}{M_i}=\Gamma_i
\left(\frac{V}{M_i} \right)^2. \label{xy} \eea
 Then, by considering  the light neutrino mass eigenvalue
equation, $ det[m_{eff} m_{eff}^{\dagger}- n^2]=0$, the lepton
number asymmetry $\epsilon_1$  and the phase $\gamma_R$ can be
written in terms of $x_1, x_2, n_2,$ and $n_3$,
\bea \epsilon_1= -\frac{3 M_1}{4 x_1 V^2}
\sqrt{\left((n_{-})^2-(x_{-})^2 \right) \left((x_{+})^2-(n_{+})^2
\right)}, \label{ep} \eea \bea \cos 2 \gamma_R &=&\frac{n_2^2 +
n_3^2 -x_1^2 -x_2^2}{2(x_1 x_2 -n_2 n_3)}, \label{gamma} \eea
where $n_{\pm}=n_3\pm n_2$ and $x_{\pm}=x_1\pm x_2$.
 There are two solutions of Eq.(\ref{gamma}) leading to negative
 $\epsilon_1$;
 $\gamma_R$ and $\gamma_R - \pi$ for $0\leq \gamma_R < \pi/2$,
 which in turn gives positive baryon number via sphaleron process.
Next, let us present the parameters, $\theta_R$, $m_2$ and $m_3$
in terms of the above 6 physical quantities. From the eigenvalue
equations for $ V_R \frac{(m_D^{\dagger} m_D)}{\sqrt{M_1 M_2}}
{V_R}^{\dagger}$ we can express $\theta_R$, $m_2$, and $m_3$ as
follows; \bea \left( m_2^2, m_3^2 \right) =\sqrt{M_1
M_2}\left(\sigma_{+}-\rho,
\sigma_{+}+ \rho \right), \label{m2m3} \\
\left(\cos\theta_R,\sin \theta_R \right)= \left(
\sqrt{\frac{\sigma_{-}+\rho}{2 \rho}},
-\sqrt{\frac{-\sigma_{-}+\rho}{2 \rho}} \right), \label{thetaR}
\eea where $ \sigma_{\pm}=\frac{x_2\pm x_1 R }{2 \sqrt{R}}$,
$\rho=\sqrt{(x_1 x_2- n_2 n_3)+ \sigma_{-}^2},$ and $R=M_1/M_2$.
We also determine $ \phi, \alpha,$ and $\theta$ with a given set
of parameters $( x_1,x_2, R, n_2, n_3)$ by using the same
procedure given in Eqs.(\ref{K},8). We take $R=0.1$.
In order to determine the values of $\phi, \theta$ and $\alpha$,
it is necessary to determine those of $x_i(i=1,2)$. Let us now
show how the variables $x_i$ can be constrained. From the neutrino
mass eigenvalue equation, it follows that \bea
&& |x_1-x_2| \le n_3 - n_2, \nn \\
&& n_3 + n_2 \le x_1 +x_2. \label{ran} \eea From the experimental
results for the neutrino oscillation, let us take $
n_3=\sqrt{\Delta m _{atm.}^2}\sim 5 \times 10^{-2}$ \mbox{eV} and
$ n_2=\sqrt{\Delta m _{solar}^2}= 7 \times 10^{-3}$ eV
(LMA)\cite{neutrino}.
From Eq.(\ref{ran}), the lower bound on $x_1$ is $0.007$ eV.
By solving the Boltzmann equation \cite{pu}, we can obtain a value
of $Y_L=\frac{n_L}{s}$, i.e., lepton number density ($ n_L $)
normalized by entropy density ($ s $). When solving the Boltzmann
equation, we need the value of $ \epsilon_1 $. For fixed $x_1$,
one can get the maximum value of $-\epsilon_1$ which gives the
maximum $-Y_L$ via Boltzmann equation. In Fig.1, we plot the
maximum lepton number density $-Y_L$ predicted from Eq.(11) as a
function of $x_1$ for several fixed $M_1$.We set the initial conditions for Boltzmann equation at $10^{16}$
GeV and we take the distribution of the heavy majorana particle
$N_1$ in thermal equilibrium and $Y_L=0$ at the temperature. 
\begin{figure}[htbp]
\begin{center}
\includegraphics[width=.7\linewidth]{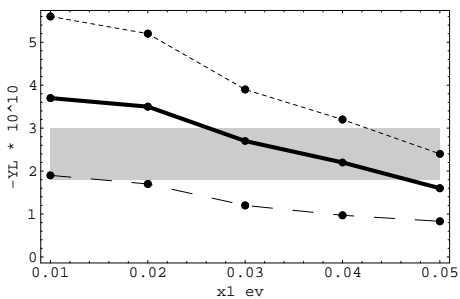}
\caption{The maximum possible lepton number density ($-Y_L$) as a
function of $x_1$  for three different $M_1$. From up to bottom,
$M_1= 3,2,1 \times 10^{11}$ \mbox{GeV}.}
\end{center}
\end{figure}
The allowed values of $-Y_L$ consistent with baryogenesis are
presented by shaded band in Fig.1. Thus, we can obtain the allowed
region of $x_1$ for a fixed $M_1$. However, there is no allowed
value of $x_1$ for a rather lower value of $M_1 < 1.0 \times
10^{11}$ GeV, which in turn leads to the lower bound on $M_1$. By
using the allowed region for $x_1$ as given in the above, we can
estimate the allowed region of $x_2$ via Eq.(11) again. Figure 2
shows how we can get the allowed region of $x_2$. For example, for
a given set $M_1=2 \times 10^{11}$ GeV, and $x_1=0.03 $ eV, we
obtain the allowed range for $x_2$ as $ 0.03  <x_2 <0.07$ eV.
\begin{figure}[htbp]
\begin{center}
\includegraphics[width=.7\linewidth]{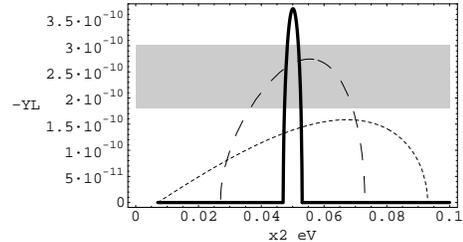}
\caption{ Lepton number density ($-Y_L$)
 as a function of $x_2$  for  $M_1= 2 \times  10^{11} $ GeV. Solid line
 corresponds to $x_1=0.01$ eV. The long dashed line corresponds to
 $x_1=0.03$ eV and the dotted line corresponds to $x_1=0.05$ eV.}
\end{center}
\end{figure}

Let us move to the other category of the parameters, $
\theta_{L12},\theta_{L23},\theta_{Ls13}, \gamma_L, \delta_L $ in
$U_L$, which are not determined from high energy phenomena, but
must be related to the low energy MNS mixing matrix, Thus two of
them can be determined from the neutrino oscillation experimental
results. For simplicity, we focus on the case with the small
mixing angles $\theta_{L13}$ and $\theta$, which is consistent
with Chooz experiment \cite{chooz}. In this case, the MNS mixing
matrix is given, in the leading order, by \bea
&& U_{MNS}\simeq \nn \\
&&\left( \begin{array}{ccc}
    c_{L12} & s_{L12} & s_{L13} e^{-i \delta_L}
    +s_{L12} s_\theta e^{-i \phi'}\\
    -s_{L12} c_{L23} & c_{L12} c_{L23} & s_{L23} \\
    s_{L12} s_{L23} & -c_{L12} s_{L23} & c_{L23}
    \end{array} \right) \nn \\
    && \times P(\alpha', -\alpha').
\label{MNS} \eea where $\phi^{\prime}=\phi+\gamma_L$,
$\alpha'=\alpha-\frac{\gamma_L}{2}$ and $
P(\alpha',-\alpha')=diag.[1, e^{i \alpha'}, e^{-i \alpha'}]$. Note
that we do not present the subleading contributions in
$(U_{MNS})_{ij}, (ij) \ne (e3)$, which are comparable to
$(U_{MNS})_{e3}$. Taking $\theta_{23} \simeq \theta_{12} \simeq
\frac{\pi}{4}$ which lead to bi-large mixing pattern, in this
approximation, $(U_{MNS})_{e3}$, $J$ and $|(m_{eff})_{ee}|$  are
given by: \bea &&|(U_{MNS)_{e3}}|\simeq |s_{L13} e^{-i \delta} +
\frac{s_\theta}{\sqrt{2}}
e^{-i \phi^{\prime}}|, \nn \\
&&J \simeq \frac{1}{4} \left( s_{L13} \sin{\delta_L} +
\frac{s_\theta}{\sqrt{2}}
\sin\phi^{\prime} \right), \nn \\
&&|(m_{eff})_{ee}| \simeq |\frac{n_2}{2} e^{4 i \alpha^{\prime}}+
n_3 (s_{L13} e^{-i \delta_L} +\frac{s_\theta}{\sqrt{2}} e^{-i
\phi^{\prime}})^2|. \label{mee} \eea In principle, we are able to
fix three unknown parameters; $\gamma_L, s_{L13}$ and $\delta_L$
once the left-hand sides of Eq.(\ref{mee}) are measured. It is
then possible to quantitatively see whether the low energy CP
violation denoted by $J$ is dominated by leptogenesis phase $\phi$
or by the CP violating phases $\gamma_L$ and $\delta_L$ which are
not related to leptogenesis. First of all, let us study the
interesting case of $ \theta_{L13}=0 $, which makes the analysis
more predictive because a CP violating phase $\delta_L$ is
simultaneously suppressed. This can be understood as the extreme
case of $ \sin \theta_{L13} \ll \sin\theta$. Interestingly enough,
this case dictates that the origin of $|(U_{MNS})_{e3}|$ may come
from the mixing angle $\theta$ which is related to heavy Majorana
neutrino sector. Jarlskog determinant is then simply given by,
\bea J= \sin 2 \theta \sin\phi^{\prime} \frac{1}{8 \sqrt{2}}.
\label{Jzero} \eea Only $\gamma_L$ is a completely  arbitrary
parameter in this case and thus we can easily investigate how $J$
can be affected by $\gamma_L$. In other word, $\gamma_L$ can be
estimated through $J$ in this case. If $\gamma_L$ is turned out to
be much smaller than $\phi$,  the measurement of CP violation in
low energy experiment may directly indicate leptogenesis.
\begin{figure}
\begin{center}
\includegraphics[width=.7\linewidth]{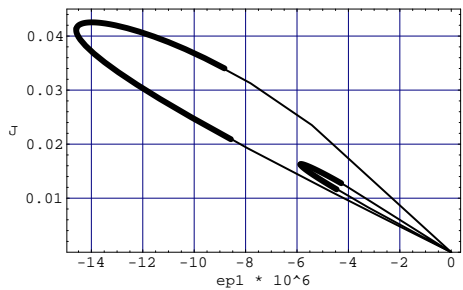}
\caption{Correlation between lepton number asymmetry $\epsilon_1$
and neutrino CP violation $J$, for $M_1=2 \times 10^{11}$ GeV. The
large (small) contour corresponds to $x_1=0.03(0.01)$ eV. The
thick solid lines show the allowed region from baryogenesis.}
\end{center}
\end{figure}
In Fig.3, we show the correlation between lepton number asymmetry
$\epsilon_1$ and $J$ in Eq.(\ref{Jzero}) with $\gamma_L=0$. In
each contour, $M_1$ and $x_1$ are fixed and $x_2$ is varied. 
For $x_1=0.01$ eV, we obtain
$|(U_{MNS})_{e3}|=\frac{\sin \theta}{\sqrt{2}} \simeq 0.066$ and
$|(m_{eff})_{ee}|=0.008$ eV, while for $x_1=0.03$ eV, $|(U_{MNS})_{e3}|$ 
is in the range $[0.15,0.2]$ and
$|(m_{eff})_{ee}|$ is in the range $[0.014, 0.018]$ eV. On the
other hand, the case with  $\sin \theta \ll \sin \theta_{L13}$, we
see from Eq.(\ref{mee}) that $J$ mainly depends on $\delta_L$,
which has nothing to do with leptogenesis.

We have studied how CP violation responsible for baryogenesis
manifests itself in MNS matrix and Jarlskog determinant which
signals low energy CP violation in neutrino oscillation.
 We have obtained cosmological constraints
on CP violation and mixing which originate from high energy
phenomena. Then using the low energy constraints we have showed it
is possible to estimate the size and sign of baryon number in the
most general case once $|(U_{MNS})_{e3}|$, neutrinoless double
beta decay and CP violation of neutrino oscillation are measured.
In a specific case of the framework, a correlation between CP
violation in neutrino oscillation and leptogenesis has been
studied and the size of $J$ has been estimated.

The works of T. M. and M. T. are supported by the Grand-in-Aid for Scientific
Research of the MEXT, Japan, No.13640290 and No.12047220 respectively. 
S.K.K is supported by JSPS Invitation Fellowship (No.L02515)
and by BK21 program of the Ministry of
Education in Korea. The authors thank H. So, K. Funakubo,
T. Kobayashi, M. Plumacher, A. Purwanto, 
KEK, YITP, and  SI2002. A part of this work was
completed during the YITP-W-02-05 on "Flavor mixing, CP violation
and Origin of matter".
%

\end{document}